%% file: plb.tex
\documentclass[aps,showpacs, twocolumn, superscriptaddress, groupedaddress,amsmath]{revtex4}

\usepackage{subfigure}
\usepackage{graphicx}
\usepackage[usenames]{color}
\usepackage{amssymb}

\newcommand{\pt}        {\mbox{$p_T$}}
\newcommand{\met}       {\mbox{$\not\!\!E_T$}}

\newcommand{\rargap}    {\mbox{ $\rightarrow$ }}
\newcommand{\ppbar}     {\mbox{$p\bar{p}$}}
\newcommand{\ttbar}     {\mbox{$t\bar{t}$}}
\newcommand{\bbbar}     {\mbox{$b\bar{b}$}}
\newcommand{\ccbar}     {\mbox{$c\bar{c}$}}
\newcommand{\comphep}   {\sc comphep}
\newcommand{\singletop} {\sc singletop}
\newcommand{\pythia}    {\sc pythia}
\newcommand{\alpgen}    {\sc alpgen}
\newcommand{\mcfm}      {\sc mcfm}
\newcommand{\geant}     {\sc geant}
\newcommand{\madgraph}  {\sc madgraph}

\begin{document}
\hspace{5.2in} \mbox{FERMILAB-PUB-13-252-E}

\title{Evidence for {\boldmath$s$}-channel single top quark production in {\boldmath$p\bar{p}$} collisions at {\boldmath$\sqrt{s}=1.96$}~TeV}

\input{author_list.tex}

\date{July 2, 2013}

\begin{abstract}
We present measurements of the cross sections for the two main production modes of single top quarks in $\ppbar$ collisions at $\sqrt{s} = 1.96$\;TeV in the Run II data collected with the D0 detector at the Fermilab Tevatron Collider, corresponding to an integrated luminosity of $9.7$\;fb$^{-1}$. The $s$-channel cross section is measured to be $\sigma({\ppbar}{\rargap}tb+X) = 1.10^{+0.33}_{-0.31}$\;pb with no assumptions on the value of the $t$-channel cross section. Similarly, the $t$-channel cross section is measured to be $\sigma({\ppbar}{\rargap}tqb+X) = 3.07^{+0.54}_{-0.49}$\;pb. We also measure the $s+t$ combined cross section as $\sigma({\ppbar}{\rargap}tb+X,~tqb+X) = 4.11^{+0.60}_{-0.55}$\;pb and set a lower limit on the CKM matrix element $|V_{tb}| > 0.92$ at $95\%$~C.L., assuming $m_t=172.5$\;GeV.
The probability to measure a cross section for the $s$ channel at the observed value or higher in the absence of signal is $1.0\times 10^{-4}$, corresponding to a significance of $3.7$ standard deviations.
\end{abstract}

\pacs{14.65.Ha; 12.15.Ji; 13.85.Qk; 12.15.Hh}

\maketitle 



\begin{figure}
\vspace{-0.1in}
\centering
\subfigure[]{\includegraphics[width=0.24\textwidth]{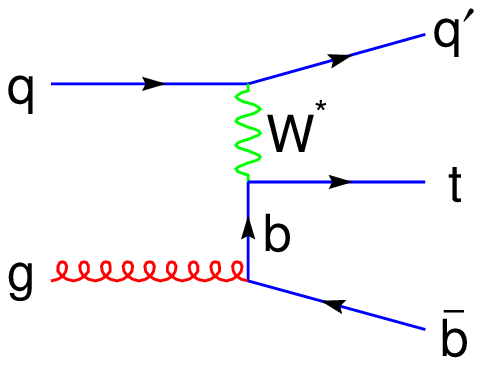}}~~
\subfigure[]{\includegraphics[width=0.24\textwidth]{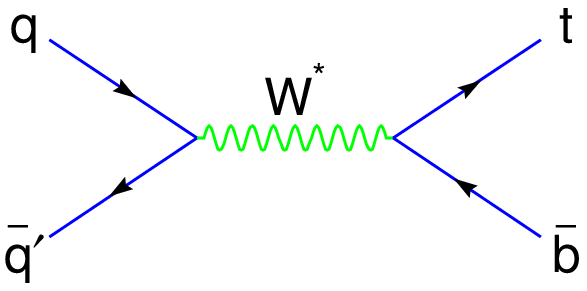}}
\vspace{-0.26in}
\caption{(Color online.) Dominant lowest-order Feynman diagrams for (a) $t$-channel and (b) $s$-channel
single top quark production.}
\label{fig:feynman_diagrams}
\end{figure}

With a mass of $m_t=173.2\pm0.9$\;GeV~\cite{topmassTeV}, the top quark is the heaviest elementary particle in the standard model (SM). The phenomenology of top quark production and decay
provides powerful means for testing the properties of the strong and electroweak interactions, as well as the possibility of discovering physics beyond the standard model (BSM) that couples strongly to mass.
At the Tevatron proton anti-proton collider operating at a center of mass energy of $\sqrt{s}=1.96$\;TeV, top quarks are produced predominantly in pairs via the strong interaction.
In addition, they are also produced by the electroweak interaction in three different production modes with a single top quark accompanied by other quarks or a $W$~boson. 
The dominant production mode at the Tevatron is the exchange of a space-like virtual $W$~boson between a light quark and a bottom quark  in the $t$ channel ($tqb = tq\bar{b}+\bar{t}qb$, where $q$ refers to a light quark or 
antiquark)~\cite{singletop-willenbrock,singletop-yuan,singletop-xsec-kidonakis}. 
The second mode is the decay of a time-like virtual $W$~boson in the $s$ channel, 
which produces a top quark and a bottom quark ($tb = t\bar{b}+\bar{t}b$)~\cite{singletop-cortese}. 
The third mode is the associated $tW$ process, in which the top quark is produced together with a $W$~boson, which contributes negligibly at the Tevatron. Figure~\ref{fig:feynman_diagrams} shows the lowest-order Feynman diagrams for the two dominant production modes at the Tevatron.

Single top quark production in the combined $s+t$ mode was observed independently by D0 and CDF in 2009~\cite{stop-obs-2009-d0,stop-obs-2009-cdf} assuming the SM ratio for the $s$- and $t$-channel cross sections. Since then, the D0 Collaboration has measured the production cross section for the combined $s+t$ channels to be $3.43^{+0.73}_{-0.74}$\;pb~\cite{stop-2011-d0}, assuming $m_t=172.5$\;GeV.
The D0 collaboration was also first to observe the $t$-channel process alone~\cite{t-channel-new} with a significance equal to 5.5 standard deviations (SD), and measured its cross section to be $2.90\pm 0.59$\;pb. 
At the CERN LHC proton-proton collider, $t$-channel production has been recently observed by the ATLAS and CMS Collaborations~\cite{atlas-tchannel,cms-tchannel}, and there is evidence for $tW$ associated production~\cite{atlas-tW,cms-tW}. Since the cross section of $s$-channel production is smaller than that of the $t$ channel and there are large irreducible backgrounds, this process has only been detected with weak statistical significance at the Tevatron, while only upper limits on the production cross section exist at the LHC. Indeed, the $s$-channel cross section increases by approximately a factor 5 from the Tevatron to the LHC at 8\;TeV, which is significantly less than the increase in the main backgrounds such as the $t$ channel, $\times 42$, or $\ttbar$, $\times 31$. The signal to background ratio is therefore better for the $s$ channel at the Tevatron.

Single top quark events can be used to 
directly measure the strength of the {\sl Wtb} vertex in the hard scatter of the collision
rather than from the decay rate as in {\ttbar} production.
The $V_{tb}$ term of the Cabibbo-Kobayashi-Maskawa (CKM)~\cite{CKM} quark-mixing matrix
is heavily constrained if one assumes that there are only three generations of quarks and that the CKM matrix is unitary, as in the SM, yielding $|V_{tb}|=0.999146^{+0.000021}_{-0.000046}$~\cite{pdg}. 
However, several BSM models predict a fourth generation of quarks or a heavy 
quark singlet that could make $|V_{tb}|$ significantly smaller than unity~\cite{Alwall:2007}. 
By measuring the rate of production of single top quark events, which is proportional to $|V_{tb}|^2$, 
this CKM element can be measured without the assumptions of three generations and unitarity of the CKM matrix~\cite{singletop-vtb-jikia}. 

In this Letter, we present improved simultaneous measurements of the $s$- and $t$-channel cross sections with the D0 Tevatron Run II  dataset corresponding to 9.7\;fb$\rm ^{-1}$ of integrated luminosity, recorded between 2002 and 2011. In addition, we provide a measurement of the $s+t$ cross section without assuming the SM ratio between the $s$ and $t$ channels. Finally, we update the measurement of $|V_{tb}|$ extracted from the $s+t$ cross section.

This analysis extends previous work by the D0 Collaboration~\cite{d0-prl-2007, d0-prd-2008, 
stop-obs-2009-d0,t-channel,t-channel-new,stop-2011-d0} and approximately doubles the integrated luminosity analyzed in the 
previous publications~\cite{t-channel-new,stop-2011-d0}. The event selection is optimized to 
maximize the $s$-channel sensitivity and to adapt to the higher instantaneous luminosity of the latest 
collected data.

Details about the D0 detector can be found in Ref.~\cite{d0-detector}.
The data are selected from an inclusive sample comprising the logical OR of many trigger conditions, which is fully efficient for the single top quark signal after offline selection. 
In the SM, top quarks decay almost exclusively to a $W$~boson and a $b$ quark. We look for leptonic decays of the $W$~boson to one electron or muon, and a neutrino. Events are therefore selected if they fulfill the following criteria:

\noindent
--There must be only one isolated electron with pseudorapidity~\cite{eta} $|\eta|<1.1$ and transverse momentum $\pt>20$\;GeV~\cite{EC} or only one isolated muon with $|\eta| < 2.0$ and $\pt > 20$\;GeV. Isolation criteria are based on calorimeter and track activity near the lepton~\cite{d0-prd-2008}. 

\noindent
--The missing transverse energy, calculated as the opposite of the vector sum of the transverse energies of all calorimeter cells surviving noise-suppression algorithms and corrected for the calorimeter energy scale and the momenta of muon tracks, is required to be $20<\met<200$\;GeV for events with two jets and $25<\met<200$\;GeV for events with three jets. The upper limit on $\met$ removes  events in data having misreconstructed muon $\pt$. 

\noindent
--We divide the sample into events requiring either two or three jets (exclusively).
All jets are required to have $|\eta| < 2.5$ and $\pt > 20$\;GeV while the leading jet
is additionally required to have $\pt > 25$\;GeV.
Jets are reconstructed by clustering cells in the calorimeter based on a cone algorithm in $(y,\phi)$ space, where $y$ is the rapidity and $\phi$ is the azimuthal angle, and the cone radius is
$0.5$~\cite{Cone}. Each jet is also required to have at least two tracks associated with the collision vertex of the $\ppbar$ hard scatter. In addition, the energy of the jet is corrected to the level of particles emitted within the jet cone~\cite{d0-jes}.

Once the basic particles in the final state are identified, we apply additional selection criteria to exclude regions of phase space that are difficult to model precisely. We require the  
scalar sum of the $\pt$ of the lepton, the $\met$, and the $\pt$ of all the jets in the event to satisfy
$H_{\rm T}({\rm jets},\ell,\met)>120$\;GeV in events with two jets, and 
$>160$\;GeV in events with three jets.
To remove multijet events where fake missing energy arises from
a jet which is misreconstructed as a lepton, we remove events having $\met$ aligned or antialigned in azimuth with the lepton by applying the following selection: $|\Delta\phi(e,\met)|>2.0-0.05\met$, $|\Delta\phi(e,\met)|>1.5-0.03\met$, $|\Delta\phi(e,\met)|<2.0-0.048\met$, and $|\Delta\phi(\mu,\met)|>1.4-0.014\met$, $|\Delta\phi(\mu,\met)|<2.5+0.021\met$.

Because signal events contain $b$ quarks, we require that one or two of the jets in each event be identified as a $b$ jet. To identify $b$ jets, a multivariate technique is used that discriminates the $b$ jets from jets produced by light quarks and gluons~\cite{D0btag}. 
Different criteria are applied to select events with one or two $b$ jets such that the 
efficiency to identify $b$ jets is $53\%$ per jet when only one $b$ tag is required, and around $65\%$ per jet when two jets are tagged in the event. The light-jet mistag probabilities in these two cases correspond on average to $0.8\%$ and $2.9\%$ per jet, respectively. The mistag probability for $c$ jets is on average less than $20\%$ per jet in the one $b$-tag channels, and $30\%$ per jet in the two $b$-tag channels.

We separate the data into four independent channels based on the 
number of reconstructed jets (two or three) and the number of $b$-tagged jets (one or at least two). 
Each of these channels has a different signal to background ratio, and by keeping them independent we improve the analysis sensitivity.
After $b$ tagging, the dominant backgrounds are $W$+jets ($63\%$ of the total background) 
and {\ttbar} events ($23\%$), which, respectively, tend to have lower $H_{\rm T}$ and larger $H_{\rm T}$ values than single top quark events.

We use Monte Carlo~(MC) generators to simulate the kinematics of signal and background events, except for the multijet events that are obtained from data.
Single top quark signal events are simulated by the {\singletop} event generator, which is based on effectively next-to-leading order (NLO) {\comphep} calculations and preserves the spin information in the decays of the top quark and the $W$~boson~\cite{singletop-mcgen}.
The simulated event kinematics match the distributions predicted by NLO 
calculations~\cite{singletop-xsec-sullivan,Campbell:2009ss}. 
The {\ttbar}, $W$+jets, and $Z$+jets events are simulated with the {\alpgen} leading-order MC 
generator~\cite{alpgen}. Diboson processes are modeled using {\pythia}~\cite{pythia}. For all these MC 
samples, {\pythia} is also used to evolve parton showers and to model proton remnants and 
hadronization of all generated 
partons. The top quark mass in single top events and {\ttbar} events is set to $m_t=172.5$\;GeV, which is within the experimental uncertainty of the current world average~\cite{topmassTeV}.
A leading-order parton distribution function, CTEQ6L1~\cite{cteq}, is used for all 
MC simulated samples, except for the $t$-channel process, which employs the NLO parton distribution 
function CTEQ6M1 to ensure the final kinematics match those calculated at NLO for that process. In addition, the factorization scale is chosen as $m_t$ for the $s$ channel, and $m_t/2$ for the $t$ channel, as prescribed in Ref.~\cite{singletop-factscale}.
The presence of additional $\ppbar$ interactions is modeled by overlaying events selected from random beam crossings matching the instantaneous luminosity profile in the data. All MC events are processed through a {\geant}-based simulation~\cite{geant} of the D0 detector, and are reconstructed using the same software as the collider data. 

Differences between simulation and data in lepton and jet reconstruction efficiencies and resolutions, 
jet energy scale, and $b$-tagging efficiencies are corrected in the simulation by applying correction 
functions  measured from separate data samples. 
The multijet background is modeled from data by selecting events that pass the 
selection described above but fail the isolation criteria for leptons. The $W$+jets and multijet backgrounds are normalized to data before $b$ tagging using the matrix method~\cite{d0-prd-2008}. All other MC 
simulated samples are normalized to the theoretical cross section at 
NNLO~\cite{ttbar-xsec} for {\ttbar}, and at NLO~\cite{mcfm} for $Z$+jets and diboson production.

Before the overall rate of $W$+jets production is normalized to data~\cite{d0-prd-2008}, the ratio of $W$+heavy flavor jets ($b$ or $c$) to $W$+light jets is set from NLO calculations~\cite{mcfm}, which correct the {\alpgen} production cross section by a factor 1.47 for $W$+$\bbbar$ and $W$+$\ccbar$ production, and by 1.65 for $W$+$c$+jets. These values are consistent with dedicated D0 measurements~\cite{whf_d0}.

\begin{figure*}
\centering
\includegraphics[width=0.325\textwidth]{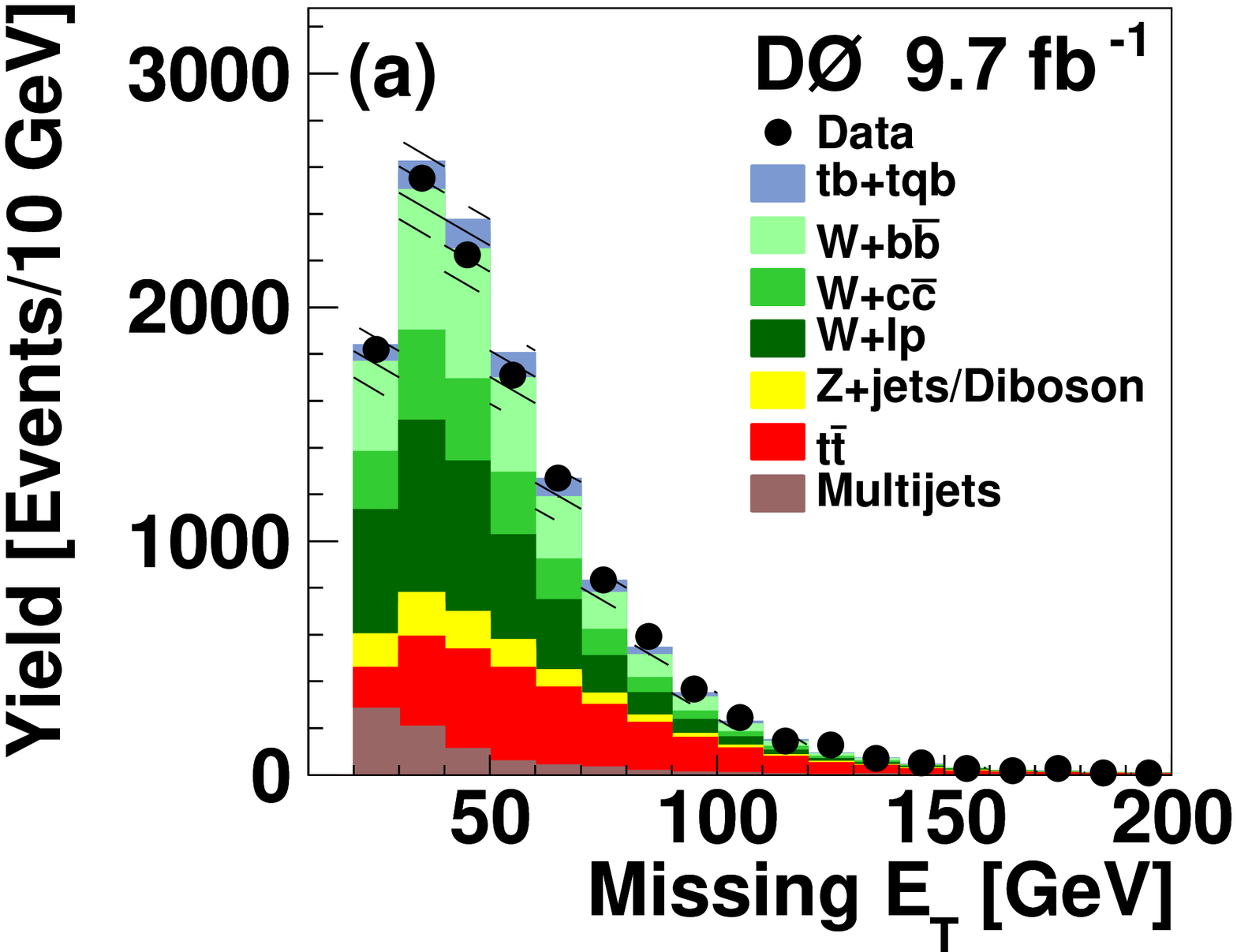}
\includegraphics[width=0.325\textwidth]{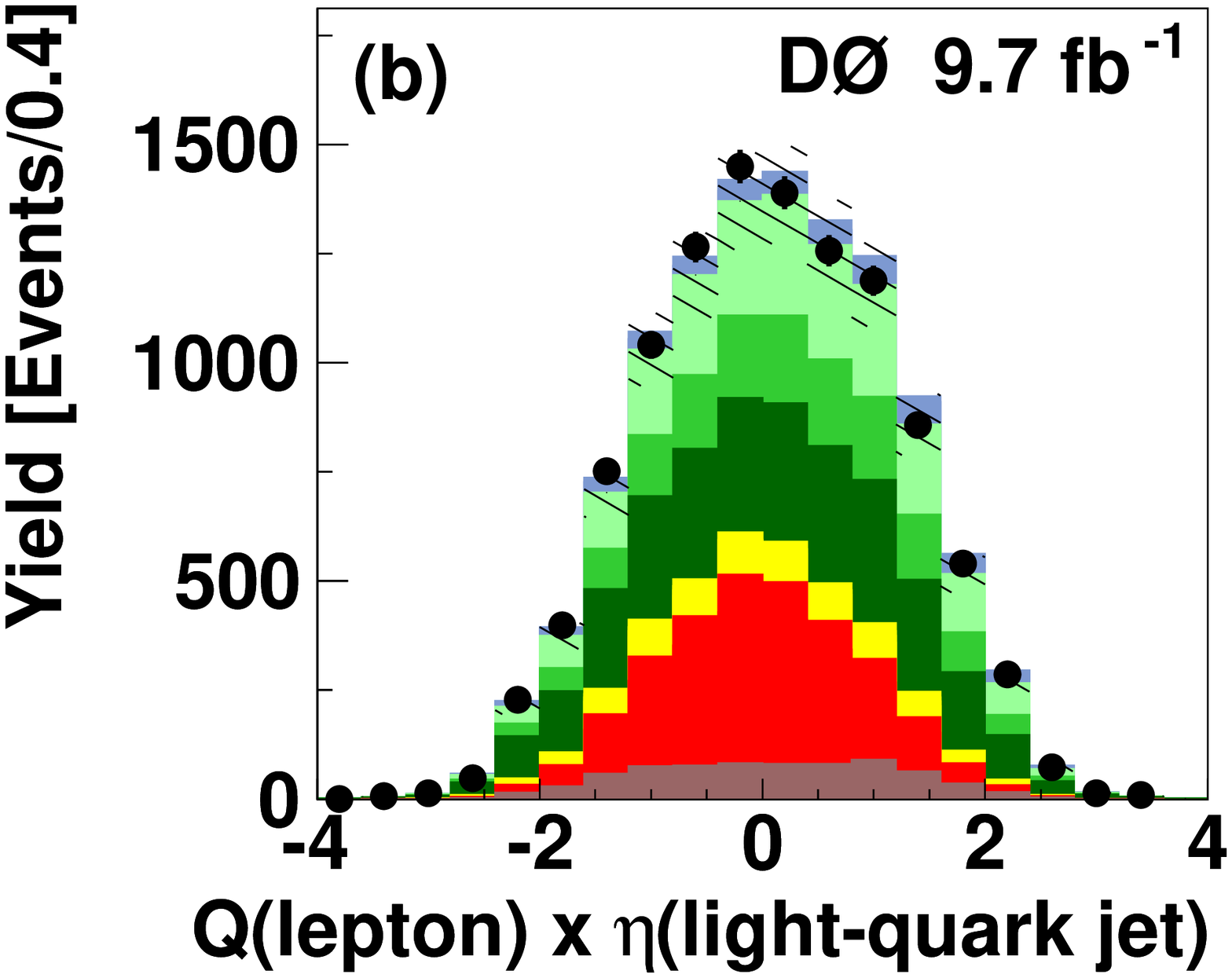}
\includegraphics[width=0.325\textwidth]{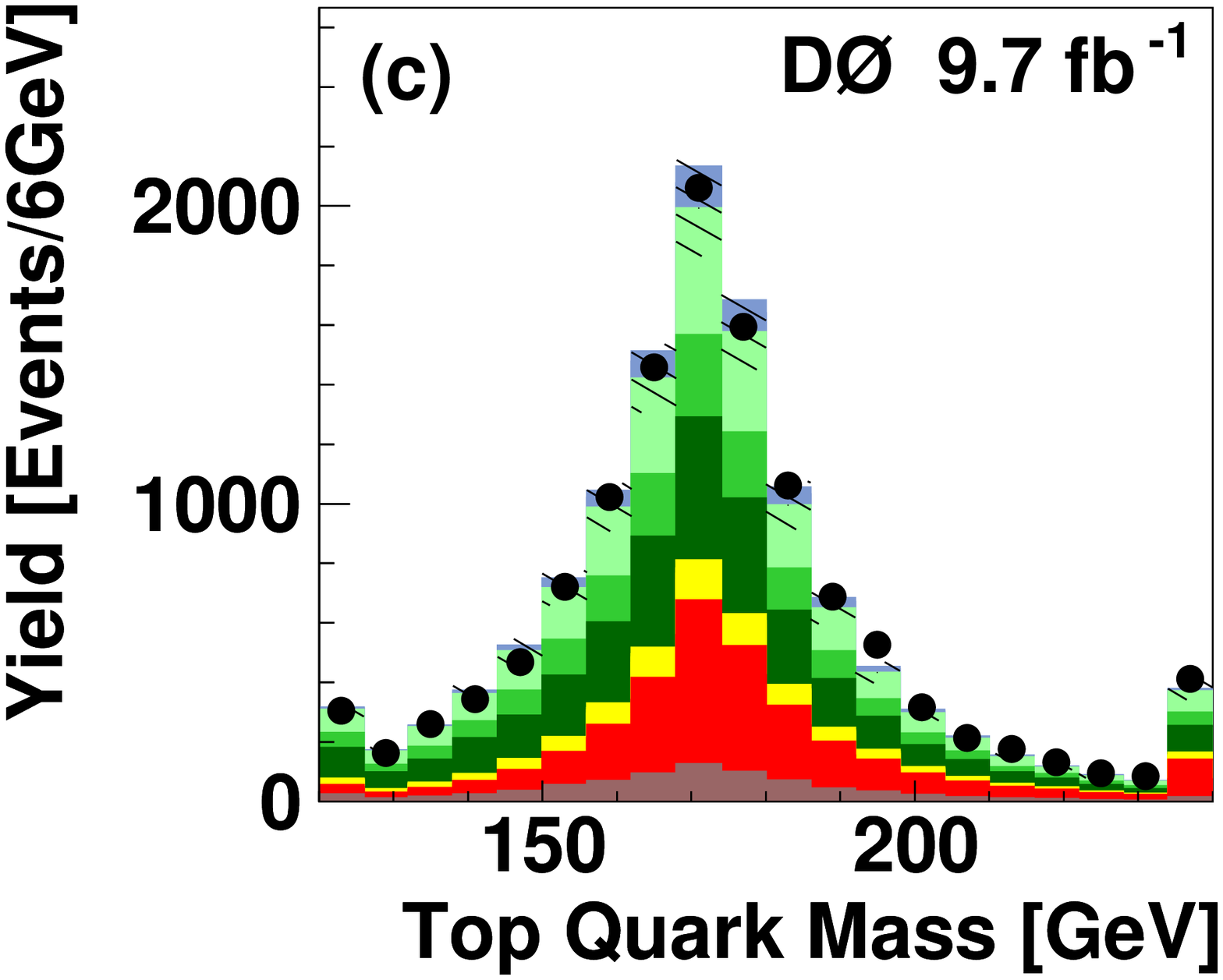}
\includegraphics[width=0.325\textwidth]{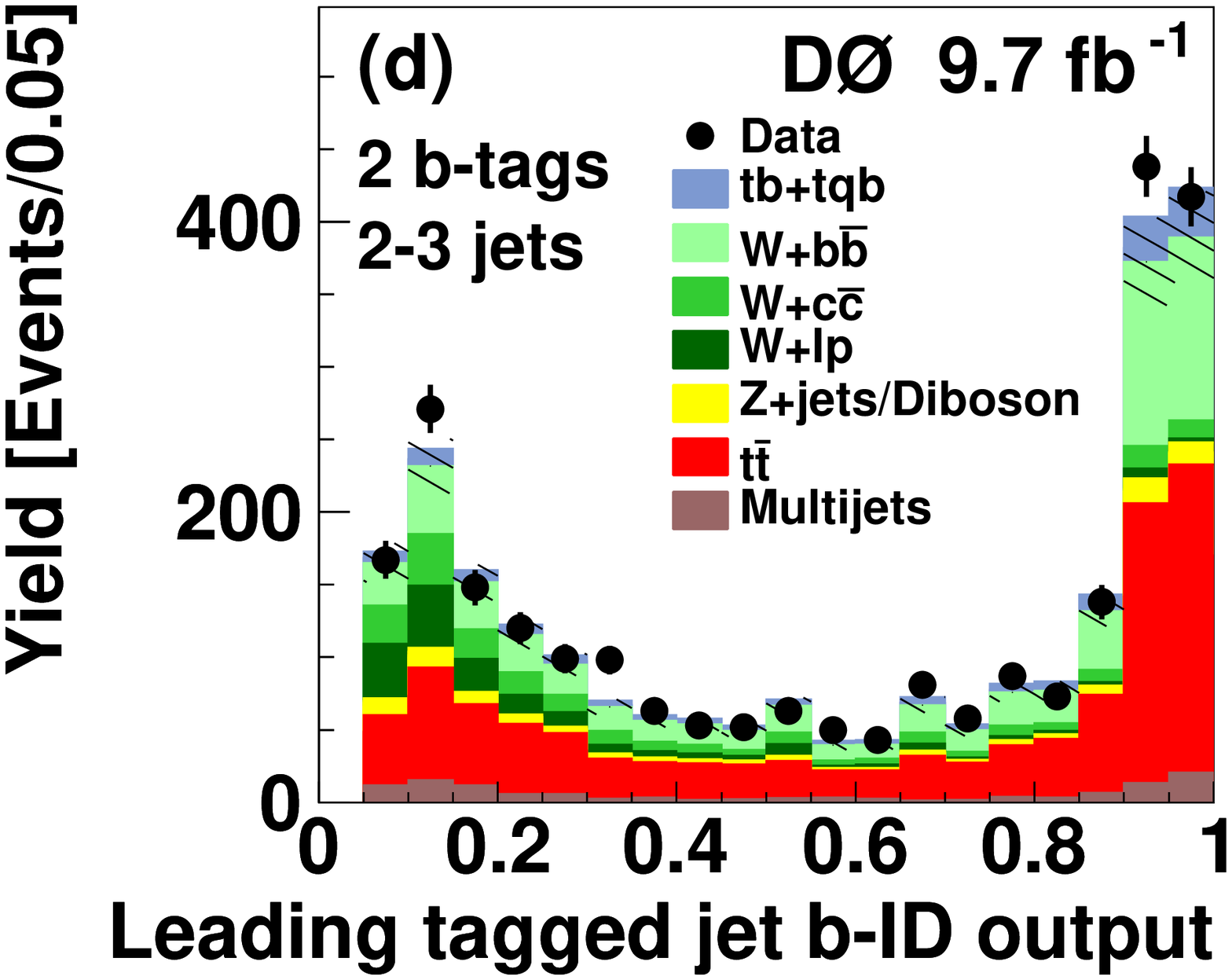}
\includegraphics[width=0.325\textwidth]{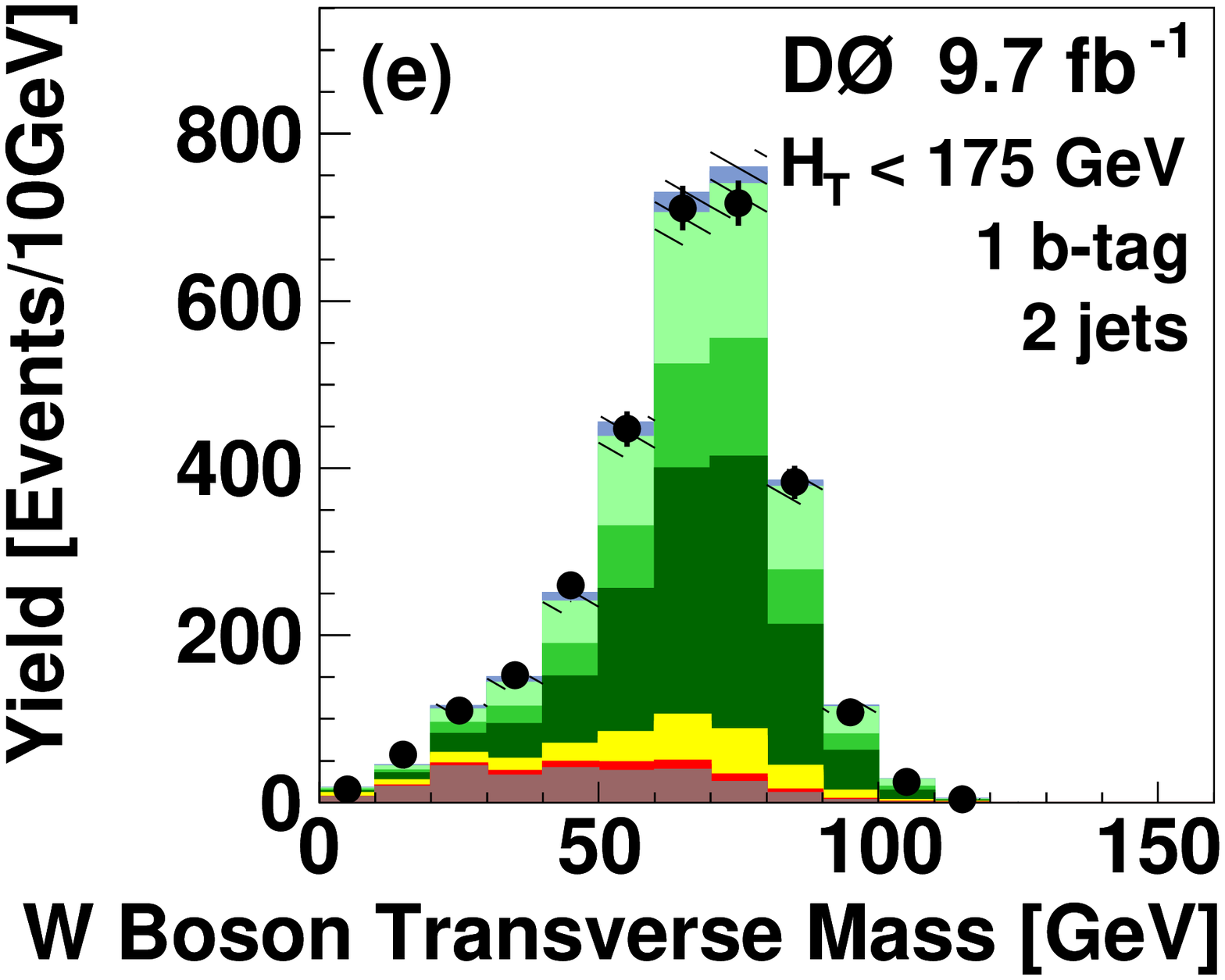}
\includegraphics[width=0.325\textwidth]{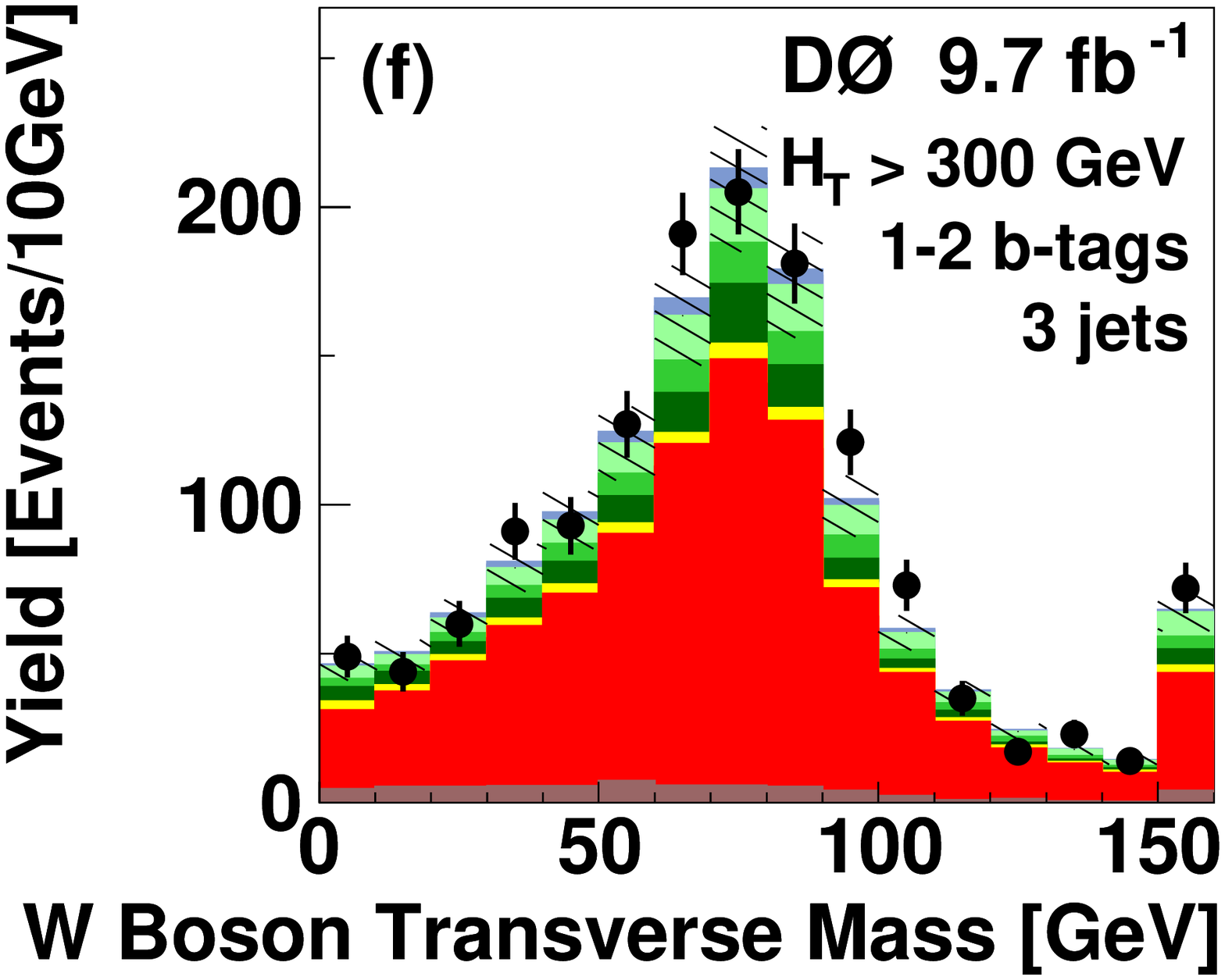}
\caption{(Color online.) 
Comparisons between the data and the background model for all channels combined for  
(a) missing transverse energy $\met$, (b) product of light quark jet pseudorapidity and lepton charge, (c) reconstructed $m_t$ from adding the $W$~boson and $b$-tagged jet that give the best fit to $m_t=172.5$\;GeV, and (d) $b$-tagging multivariate output for the most $b$-like jet after the heavy flavor correction in the two $b$-tag channel. The selection threshold for one tagged jet is set at 0.225, and for two $b$ jets is set at 0.075 each in this distribution. Also shown is the $W$~boson transverse mass (e) in a control sample dominated by $W$+jets, and (f) in a control sample dominated by {\ttbar} pairs. The $s$- and $t$-channel contributions are normalized to their SM expectations for $m_t=172.5$\;GeV. Underflow and overflow entries are added to the first and last bins, respectively. The hatched bands show the $\pm 1$\;SD uncertainty on the background prediction.}
\label{fig:variables}
\end{figure*}

To properly describe the kinematics of a $W$+jets enriched sample and of the $W$+jets dominated region, we renormalize the simulation in the last two bins of the $b$-tagging multivariate output (0.90,0.95) and (0.95,1.00). The correction factor is derived from a sample that has low values of the matrix element discriminant, described below. These events are highly depleted in signal (signal fraction $<$1\% after $b$ tagging), and are dominated by $W$+jets production in the two-jet channel. They provide enough statistics to derive a correction factor valid also for the highest bins of the $b$-tagging multivariate output, as shown in Fig.~\ref{fig:variables}(d). These events are not used in the subsequent measurements. This correction scales down the simulated samples by an average factor of $0.80 \pm 0.08$, where the uncertainty is statistical only.
The total uncertainty assigned to this normalization is $20\%$, which is consistent 
with studies in independent data sets with no $b$-tagged jets, and with fits of data to the 
$b$-tagging output of the background components in different channels. 

Table~\ref{tab:event-yields} lists the numbers of expected and observed events for each process after 
event selection including $b$ tagging. Overall, the total combined acceptance including the branching fraction, event selection, 
and $b$ tagging, is $2.6\%$ for $s$ channel and $1.8\%$ for $t$ channel. 
\begin{table*}[!h!tbp]
\vspace{-0.15in}
\begin{center}
\caption{The numbers of expected and observed events in a data sample corresponding to 9.7\;fb$\rm ^{-1}$ of integrated luminosity, with uncertainties including both statistical and systematic components added in quadrature,
before the fit to the data. The $s$- and $t$-channel contributions are normalized to their SM expectations for $m_t=172.5$\;GeV. 
The ratio $S(tb)$:$B$ is the ratio of the number of $s$-channel signal events, $S$, to the 
total number of background events, $B$, including the $t$ channel, and $S(tqb)$:$B$ is the ratio of the number 
of $t$-channel signal events to the total number of background events, including the $s$ channel.}
\label{tab:event-yields}
\begin{ruledtabular}
\begin{tabular}{l@{\extracolsep{\fill}}r@{\extracolsep{0pt}$\pm$}l@{}%
                 @{\extracolsep{\fill}}r@{\extracolsep{0pt}$\pm$}l@{}%
                 @{\extracolsep{\fill}}r@{\extracolsep{0pt}$\pm$}l@{}%
                 @{\extracolsep{\fill}}r@{\extracolsep{0pt}$\pm$}l@{}}
Number of jets     & \multicolumn{2}{c}{~2} & \multicolumn{2}{c}{~2} & \multicolumn{2}{c}{~3} & \multicolumn{2}{c}{~3}\\
Number of $b$ tags & \multicolumn{2}{c}{~1} & \multicolumn{2}{c}{~2} & \multicolumn{2}{c}{~1} & \multicolumn{2}{c}{~2}\\ \hline
$s$ channel           & 112 & 23  & 83   & 19  & 33   & 7   & 29  & 7  \\
$t$ channel           & 248 & 50  & 23   & 5   & 75   & 15  & 32  & 7  \\
{\ttbar}              & 585 & 100 & 275  & 52  & 1044 & 207 & 767 & 158  \\
$W$+jets              & 4984& 369 & 715  & 96  & 1395 & 120 & 300 & 39  \\
$Z$+jets and diboson  & 544 & 67  & 79   & 10  & 156  & 18  & 36  & 5  \\
Multijet              & 479 & 73  & 65   & 10  & 188  & 33  & 56  & 9  \\ \hline
Background sum        & 6592 & 395  & 1134 & 110 & 2784 & 242  & 1160 & 164  \\
Backgrounds + signals & 6952 & 399  & 1240 & 112 & 2891 & 243  & 1220 & 164  \\ \hline
Data                  & \multicolumn{2}{c}{~6859} & \multicolumn{2}{c}{~1286} & \multicolumn{2}{c}{~2725} & \multicolumn{2}{c}{~1233}  \\ \hline
$S(tb)$:$B$           & \multicolumn{2}{c}{~1:61} & \multicolumn{2}{c}{~1:14} & \multicolumn{2}{c}{~1:88} & \multicolumn{2}{c}{~1:41}  \\
$S(tqb)$:$B$          & \multicolumn{2}{c}{~1:27} & \multicolumn{2}{c}{~1:52} & \multicolumn{2}{c}{~1:38} & \multicolumn{2}{c}{~1:38} \\
\end{tabular}
\end{ruledtabular}
\end{center}
\vspace{-0.15in}
\end{table*}
 
Figure~\ref{fig:variables} shows comparisons between data and 
simulation after applying $b$ tagging, with all corrections included. In the same figure, the 
normalization and differential spectra of the two dominant backgrounds are checked
using the control samples dominated by $W$+jets events (Fig.~\ref{fig:variables}(e)), and by {\ttbar} events (Fig.~~\ref{fig:variables}(f)).
These plots demonstrate the accuracy of the background modeling.

There is no single kinematic variable that allows for the efficient isolation of the single top quark 
signal from the large backgrounds. 
We therefore perform the following three separate multivariate analyses (MVA) and then combine their results into one final MVA:
(i) matrix element (ME)~\cite{me-method}, 
(ii) Bayesian neural networks (BNN)~\cite{bayesianNNs}, and
(iii) boosted decision trees (BDT)~\cite{decision-trees}. The final combination of these three separation techniques is performed using a BNN.
By combining several input variables, each method defines a 
discriminant output variable $D$ between 0 and 1 where the signal tends to be in the high discriminant 
region ($D \approx 1$). The output $D$ achieves better signal separation than any single kinematic variable and is used to extract the signal in the high discriminant region by fitting the data to the sum of the signal and background models, with the signal and background normalization as free parameters. The background normalization is thus constrained by the data with low discriminant values. 

For all three MVA methods, we use the same data and the same model for background, perform the analyses 
separately on the four mutually exclusive channels defined previously, and consider the same sources 
of systematic uncertainty. All MVA methods produce discriminants for the $s$ channel as signal, $D_s$, 
where the $t$ channel is treated as another background, and for the $t$-channel signal, $D_t$, 
where the $s$ channel is considered in the background category. The three methods differ, however, in the discriminating variables. 

The ME technique uses the theory that governs the production of signal and background events to separate them. The matrix element $\mathcal{M}$ for a given process contains all the dynamics of the hard scattering, where a collision between two initial partons $p_1$ and $p_2$ produces the final state partons described by their four-momenta $k$. Thus, the differential cross section for a given process $p_1p_2\to k$ is proportional to the magnitude squared of the ME for that process: $d\sigma / dk \propto |\mathcal{M}(p_1p_2\to k)|^2$.
The ME method uses the probabilities derived from these differential cross sections to create a discriminant that potentially uses all the kinematic information available for the event. In our background probability calculation, we include the MEs for all backgrounds, including multijets, as described by the dominant leading-order Feynman diagrams obtained from {\madgraph}~\cite{madgraph}. 

Additional details about the specific implementation of the ME method for this analysis can be found in Ref.~\cite{ME-thesis}.
In this analysis, we have improved several aspects of the method with respect to the previous implementation~\cite{d0-prd-2008,t-channel}:

\noindent
--The {\ttbar} process produces a six parton final state: $\ell^{\pm}\nu b q\bar{q}'\bar{b}$ or $\ell^+\nu b \ell^-\bar{\nu}\bar{b}$, but the analyzed final state contains at most five partons. We could integrate over the phase space of the extra partons in each event, but we instead choose to match each parton to a reconstructed object in our final state to speed up the calculations. We find that the missing jets are most frequently light quark jets originating from the $W$~boson decay. 
In the two-jet channel the $W$ boson decaying hadronically is therefore assumed to be lost and is integrated over with a prior obtained from a simulation of {\ttbar} events with two reconstructed jets. In the three-jet channel, the optimal procedure is to assign the $W$~boson momentum before it decays to the third (light) jet with a corresponding transfer function that takes into account the average energy carried away by the lost jet.

\noindent
--The transfer functions that relate the reconstructed jet energy to the parton-level energy have been updated to provide improved modeling of energy resolutions. We treat jets misidentified as electrons, light jets, $b$ jets, and $b$ jets with muon decays inside the jet as separate categories for each jet transfer function.

\noindent 
--New discriminants have been introduced that incorporate the $b$-tagging information for each jet into the ME probabilities to improve the characterization of each event. In the $t$-channel discriminant each jet-parton permutation is assigned a weight based on the $b$-tagging output of the jet. In the $s$-channel discriminant all jet-parton permutations have equal weights. The overall probability is increased if the $b$-tagging information of the jets in the event matches the expected number of $b$ jets for each ME process. In this case, the added $b$-tagging information helps in discriminating the signals from backgrounds that contain light jets.

The BNN and BDT methods are different from the ME method because they rely on the simulated samples to characterize the signals and backgrounds, instead of using the ME for each process. The BNN and BDT follow the procedure established in the previous measurement~\cite{t-channel-new}. The selected sample is divided into three different subsamples: a quarter of the events is used for the training sample used to characterize the signal and background distributions in the BNN and BDT; a quarter is set aside for the training of the combination method (which will later combine the ME, BNN and BDT results); and the remaining half is used to check the convergence, measure the cross sections, and display the distributions of all variables. A more detailed description of this analysis is given in Refs.~\cite{bnn-thesis,bdt-thesis}.

A neural network is based on a set of non-linear functions that approximate a real function of one or more variables. Neural networks are trained to approximate the optimal discriminant that separates the signal from the background. We use a Bayesian approach to scan over many different neural networks to find the best discriminant~\cite{bayesianNNs}. The optimal neural network is found by averaging over the parameters that define each neural network, and by assigning a probability to each configuration~\cite{bnn-thesis}.
The BNN uses the momentum of the lepton, $\met$, and the momenta of the jets as input variables. For each jet, the $b$-tag multivariate output is also used. In addition, two variables are added that improve the performance of the discriminant: the transverse mass of the $W$~boson, reconstructed from the lepton and the $\met$, and the product of the leading untagged jet $\eta$ and the lepton charge, $Q(\ell)\times\eta(q)$, which characterizes the forward production in the $t$ channel. For the channel with two jets and two $b$ tags, this variable is not used. In total, the BNN uses 14 variables in the two-jet channel, and 18 variables in the three-jet channel. 

Decision trees classify events by sequentially applying selection criteria leading to several disjoint subsets of events, each with different signal purity~\cite{decision-trees}. The decision tree is built by creating two branches for the most optimal selection criterion amongst the list of input variables for the given data, and repeating this procedure with each subsequent subset.
``Boosting'' is the retraining of a previous decision tree by increasing the weight of those events that are misclassified in the parent tree, such that the new tree will focus more on signal events with low discriminant values and background events with high discriminant values.
The input variables to use for the BDT are selected by ranking a large set of well modeled variables in order of separation power optimized for the $s$-channel signal for all channels combined, and then selecting the best 30 variables. To ensure a well behaved discriminant, we only use input variables that have good agreement between data and simulation, as checked in the training sample, i.e. having a binned Kolmogorov-Smirnov test value higher than 0.25. 

\begin{figure*}[!ht]
\centering
\includegraphics[width=0.49\textwidth]{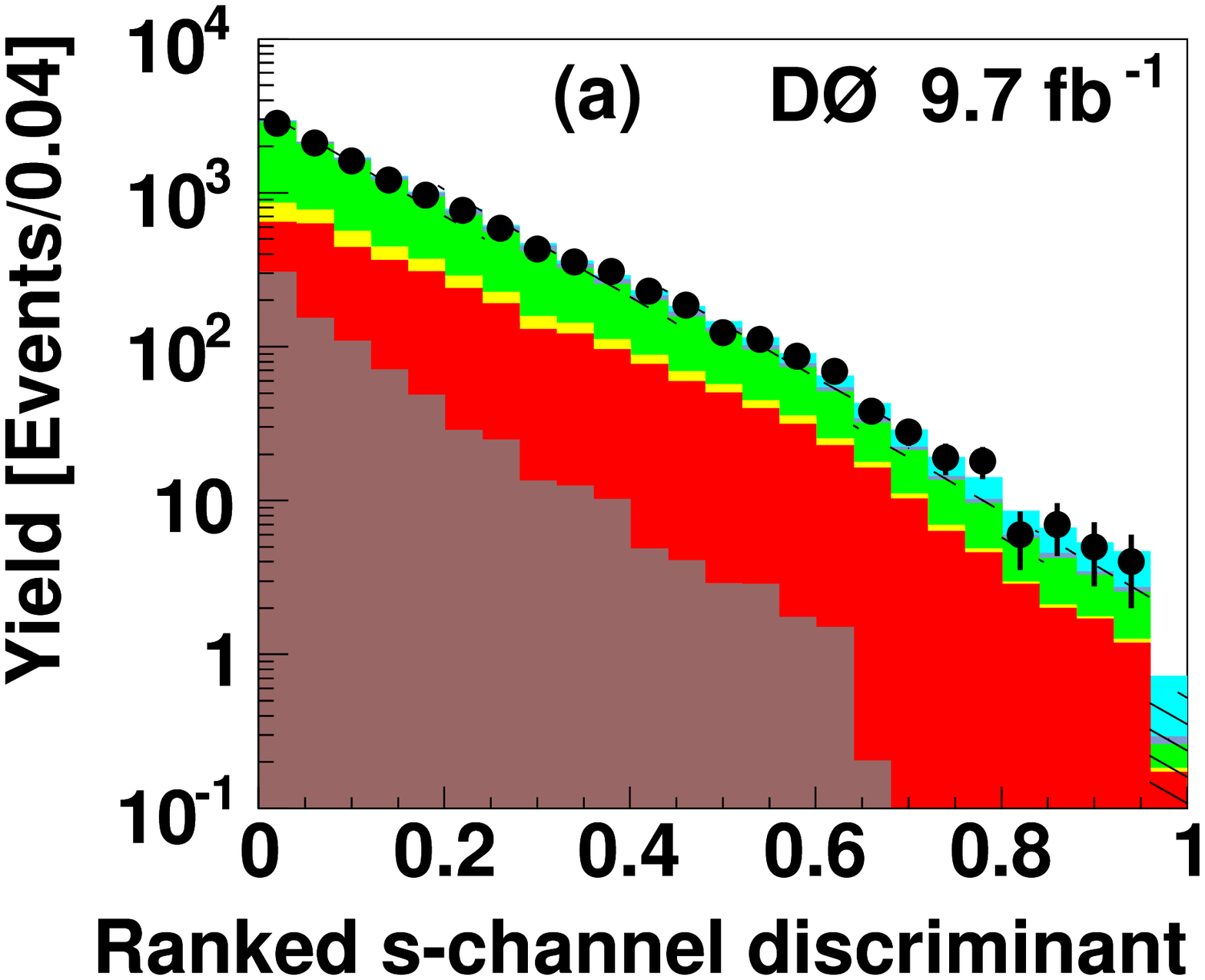}
\includegraphics[width=0.49\textwidth]{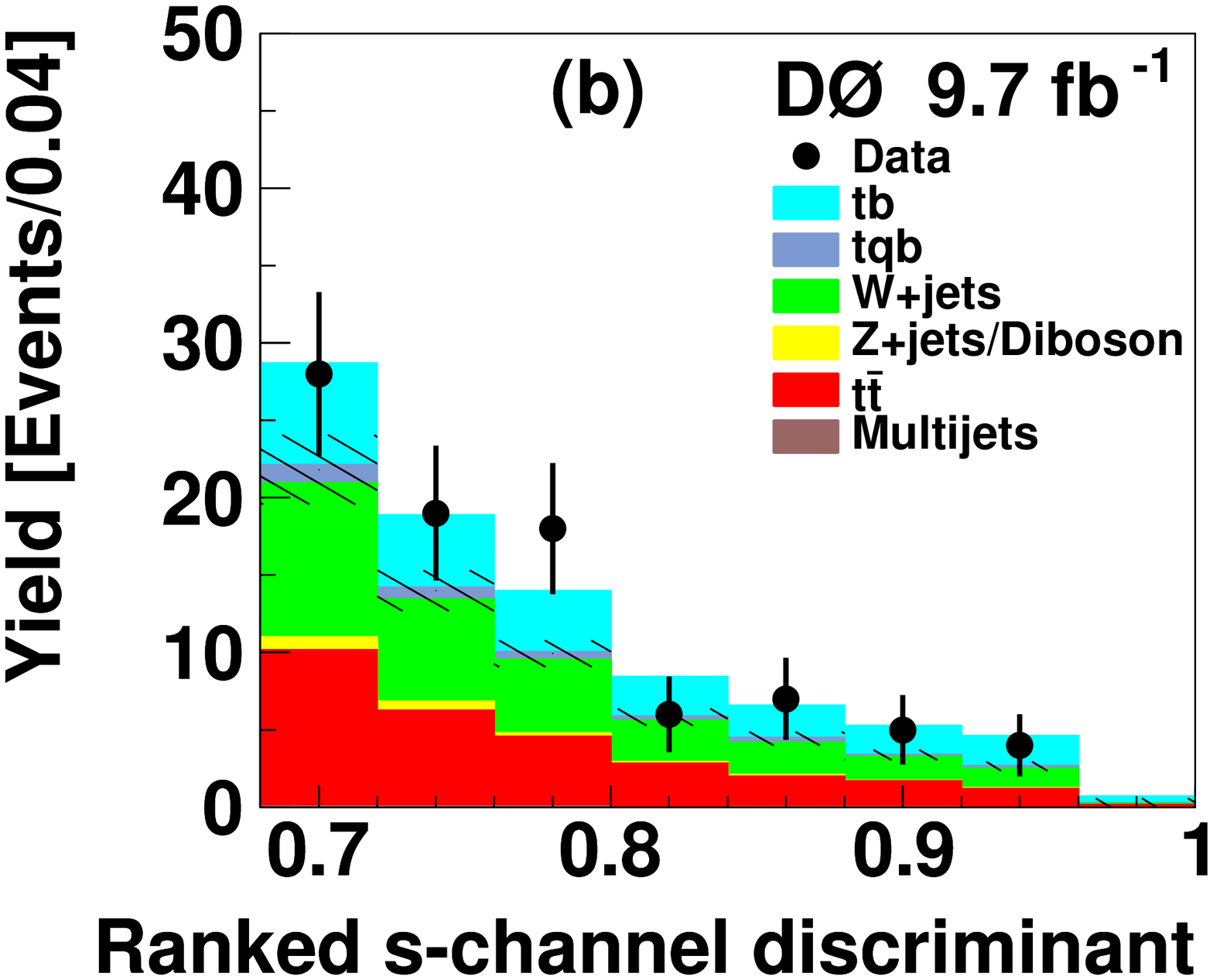}
\includegraphics[width=0.49\textwidth]{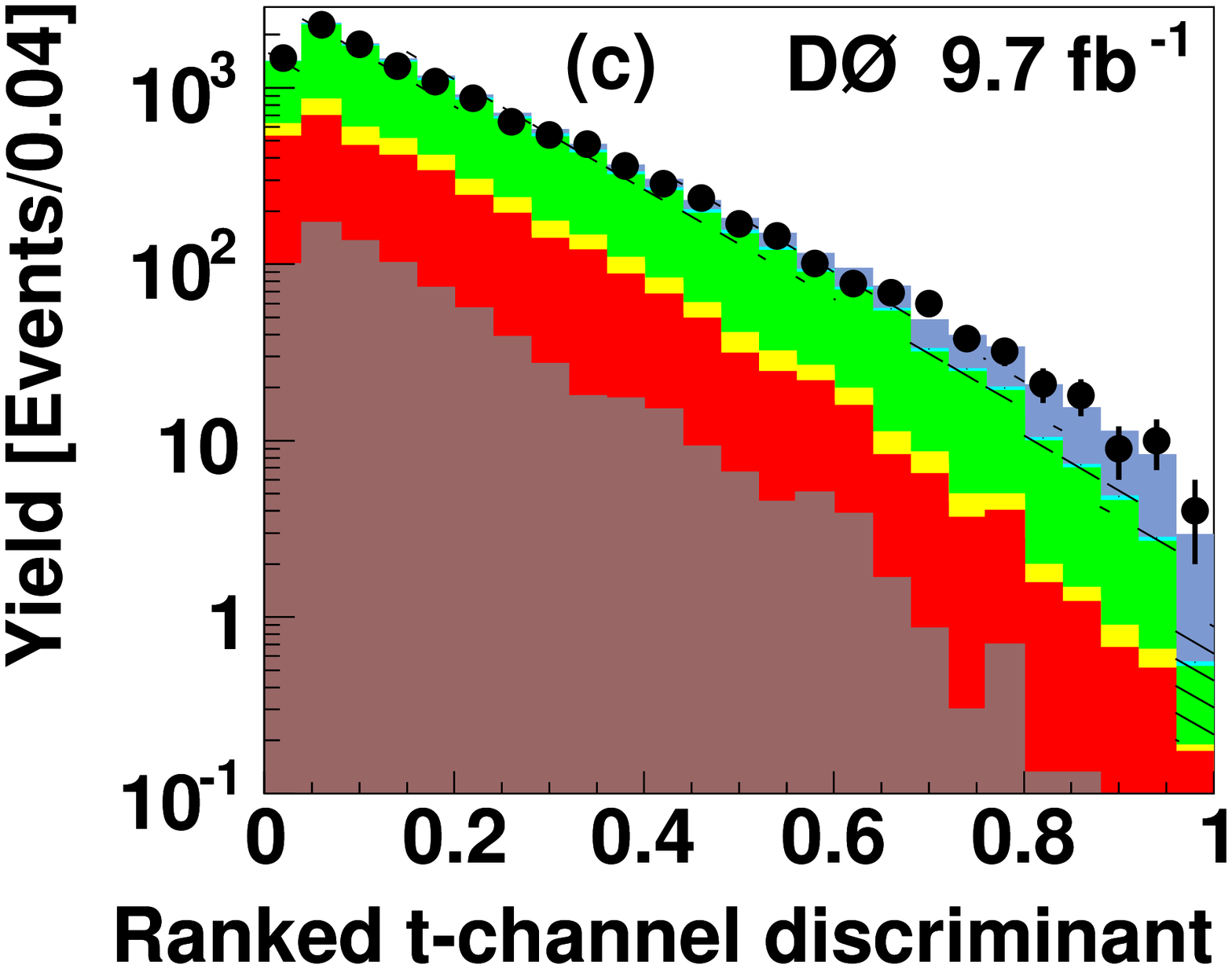}
\includegraphics[width=0.49\textwidth]{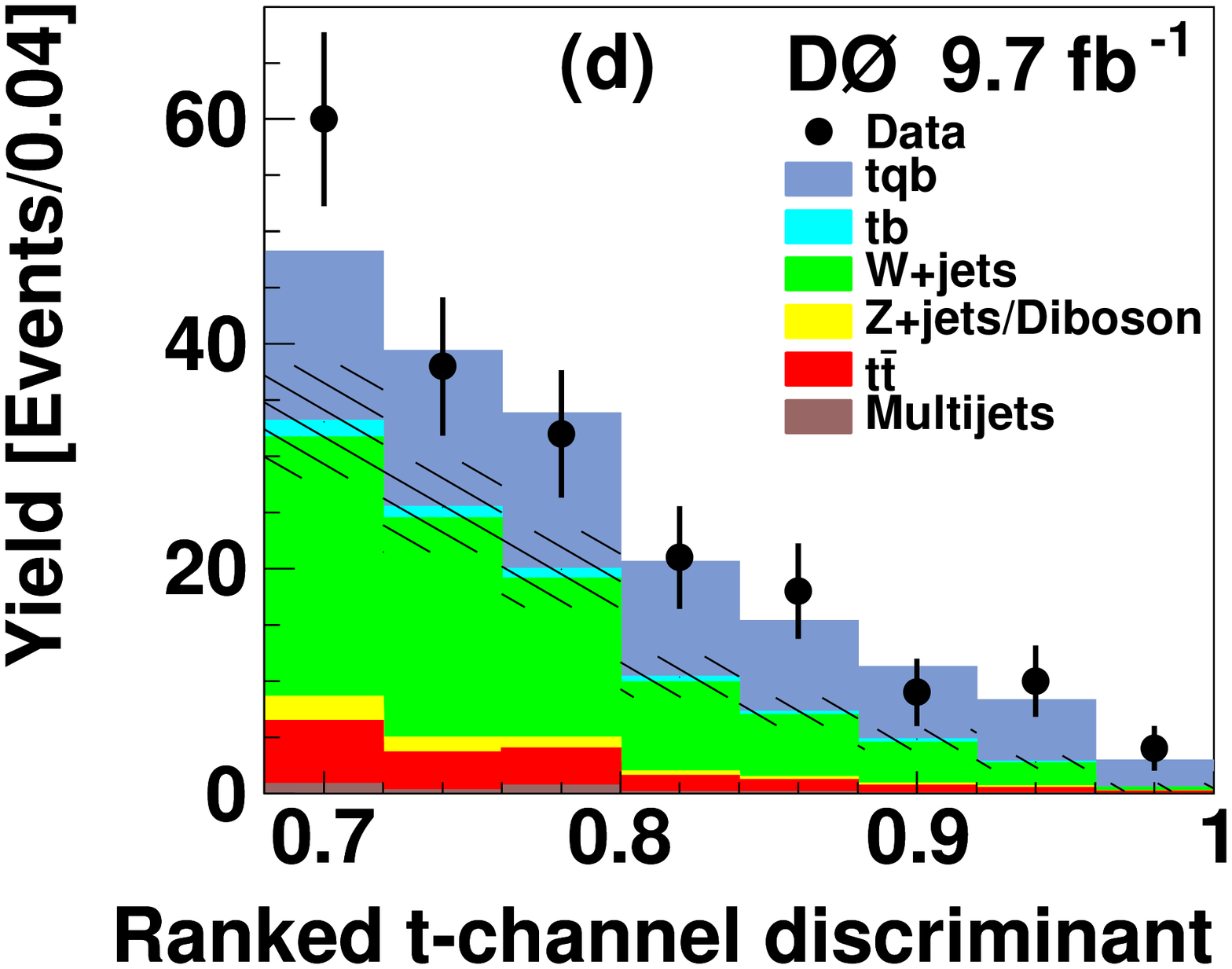}
\vspace{-0.1in}
\caption{(Color online.) The (a) $D_s^{\rm comb}$ and (c) $D_t^{\rm comb}$ discriminants for all analysis channels combined, with the high discriminant region shown in (b) and (d), respectively. The bins
have been ranked by their expected signal to background ratio. The signal is normalized to the observed cross section. The signal contributions are visible above the hatched bands that show the $\pm 1$\;SD uncertainty on the background prediction after the fit to the data.}
\label{fig:BNNCOMBdisc}
\end{figure*}

All three MVAs achieve similar discrimination between signal and background events, and their discriminants show good agreement of the background expectation with the data in the background dominated regions. 
Using ensembles of simulated datasets containing contributions from background and signal, we infer that the pairwise correlations among the outputs of the individual MVA methods are $\approx$\;75\%. 
Sensitivity can therefore be increased by combining the methods to form a new discriminant~\cite{t-channel}. 
To achieve maximum sensitivity, a second BNN is used to construct a combined discriminant for $s$- and $t$-channel signals, defined as $D_s^{\rm comb}$ and $D_t^{\rm comb}$, for each analysis channel.
The new BNN takes as input variables the three discriminants of ME, BNN, and BDT methods for the corresponding signal, and is trained on the remaining, independent, quarter of the selected sample.
Figure~\ref{fig:BNNCOMBdisc} shows that the $D_s^{\rm comb}$ and $D_t^{\rm comb}$ distributions display
agreement between the data and the expected background plus measured signal over the entire discriminant range. 

Systematic uncertainties are categorized in two classes: one only affecting the overall normalization,
and the other affecting both the normalization and the kinematic distributions and therefore the discriminant distributions. 
Table~\ref{table:sys_summary} provides a summary of the systematic uncertainties. The most important ones are due to the $W$/$Z$+jets heavy flavor corrections, which include uncertainties on the NLO scaling, and on the correction applied to the $b$-tag discriminant from the control sample; the $b$-tagging efficiency uncertainty and scale factors; and the uncertainties on some of the cross sections for backgrounds.
\begin{table}[!h!tbp]
\begin{center}
\caption[gensys]{A summary of the dominant relative systematic uncertainties.
The uncertainties shown correspond to the overall change in the yield of
the relevant signal or background components for each uncertainty source, and
are not the uncertainties on the final cross section. 
Ranges are given to cover the spread across different channels.}
{\begin{tabular}{lc}
\hline
\hline
\vspace{-0.4cm} \\
\multicolumn{2}{c}{\bf Relative Systematic Uncertainties} \vspace{0.1cm}\\
\hline
\vspace{-0.4cm} \\
{\bf Components for Normalization}               &              \\
~~Integrated luminosity~\cite{lumi}              &  6.1\% \\
~~{\ttbar} cross section                         &  9.0\%       \\
~~Parton distribution functions                  &  2.0\%       \\
~~Trigger efficiency                             &  (3.0-5.0)\% \\
~~Jet fragmentation and higher-order effects     &  (0.7-7.0)\% \\
~~Initial and final state radiation              &  (0.8-10.9)\% \\
~~$W/Z$+jets heavy flavor correction             &  20.0\%     \\
~~$W$+jets normalization to data                 &  (1.1-2.5)\% \\
~~Multijet normalization to data                 &  (9.2-42.1)\%   \\
\hline
\vspace{-0.4cm} \\
{\bf Components for Normalization and Shape}     &         \\
~~Jet reconstruction and identification          &  (0.1-1.4)\%  \\
~~Jet energy resolution                          &  (0.3-1.1)\%  \\
~~Jet energy scale                               &  (0.1-1.2)\%  \\
~~Flavor-dependent jet energy scale              &  (0.1-1.3)\%  \\
~~$b$ tagging, single-tagged                     &  (1.0-6.6)\%  \\
~~$b$ tagging, double-tagged                     &  (7.3-8.8)\%  \\
\hline
\hline
\end{tabular}}
\label{table:sys_summary}
\end{center}
\end{table}

We use a Bayesian approach~\cite{d0-prl-2007, d0-prd-2008, stop-obs-2009-d0} to extract the production cross sections. The method consists of forming a binned likelihood as a product of all four analysis channels (two or three jets with one or two $b$ tags) on the bins of the full discriminant distributions. We use the two discriminants $D_s^{\rm comb}$ and $D_t^{\rm comb}$ simultaneously in a joint discriminant sensitive to both signals, which makes the measurements of the single top quark cross sections $\sigma_s$ and $\sigma_t$ correlated.  
We assume a Poisson distribution for the number of events in each bin and uniform prior probabilities for positive values of the signal cross sections. Systematic uncertainties and their correlations are taken into 
account by integrating over signal acceptances, background yields, and integrated luminosity, assuming a 
Gaussian prior for each source of systematic uncertainty. 
A two-dimensional (2D) posterior probability density is constructed as a function of 
$\sigma_s$ and $\sigma_t$, with the position of the maximum defining the value of the cross sections, and the width of the distribution in the minimal region that encompasses $68\%$ of the entire area defining the uncertainty (statistical and systematic  components combined). The expected cross sections are obtained by setting the number of data events in each channel equal to the  value given by the prediction of SM signal plus background. 

Several cross checks have been performed to demonstrate the stability of the MVA methods and the Bayesian extraction of the cross section, and to ensure the reliability of the measurements.
We generate ensembles of pseudo-experiments taking into account all systematic uncertainties and their correlations, injected with varying amounts of signal events. Each pseudo-experiment is analyzed with each of the MVA methods, following the same analysis chain as for the data, and the signal cross section is extracted. The cross sections extracted by all three methods behave linearly as a function of the input signal cross section. The same behavior is found for the combination BNN. Results of these pseudo-experiments demonstrate insignificant biases.
We test the MVA methods in the two cross-check regions in the data, enriched in $W$+jets and {\ttbar} events, and the discriminants show good agreement with the background expectation in these background dominated samples. 
Finally, we also check the distribution of the data sample when different regions of the discriminants are selected with increasing amounts of signal purity, and show that the presence of a single top quark signal is needed to ensure a good description of the data in different kinematic variables.

\begin{figure*}[htbp]
\begin{center}
\includegraphics[width=0.49\textwidth]{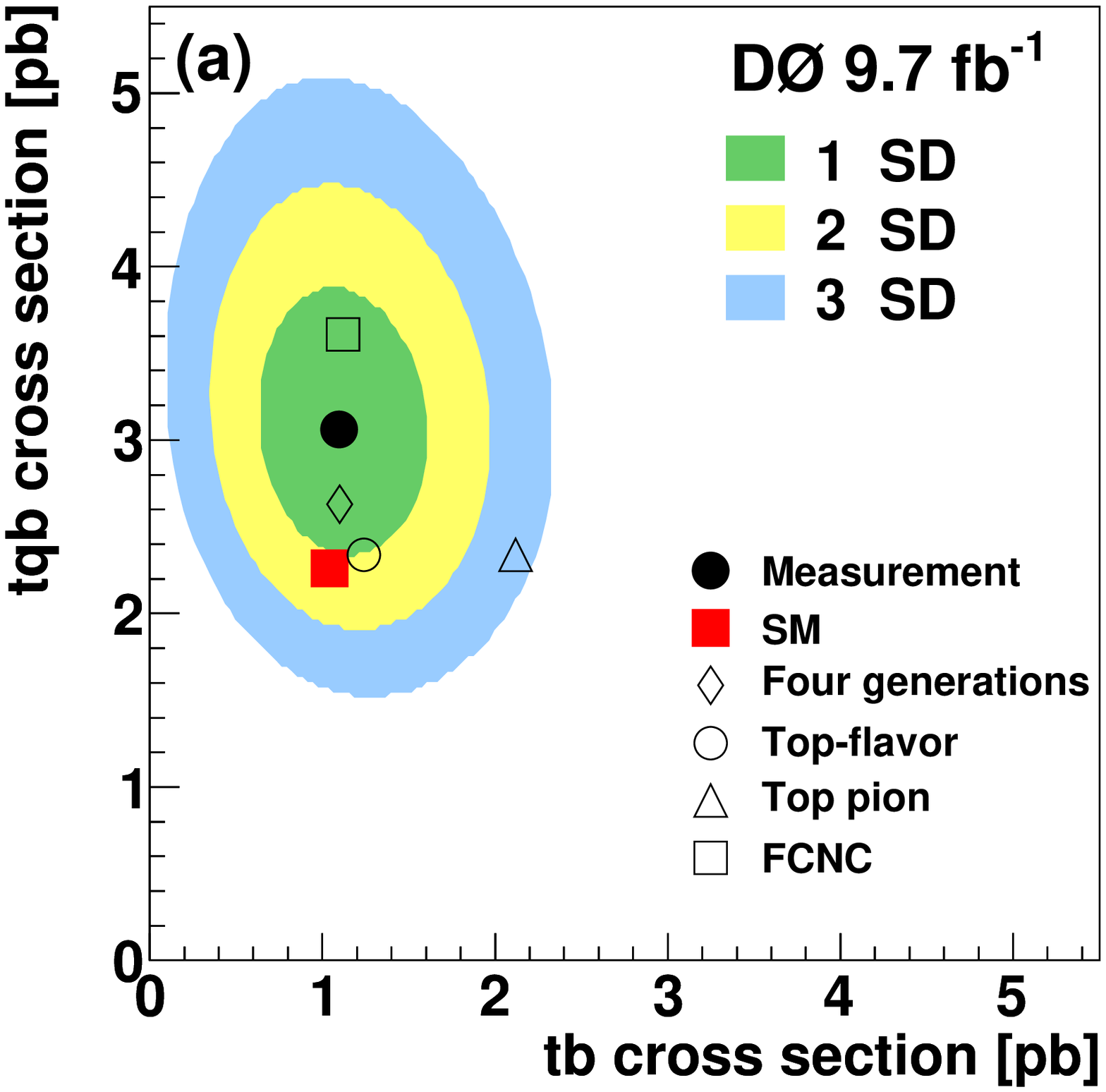}
\includegraphics[width=0.49\textwidth]{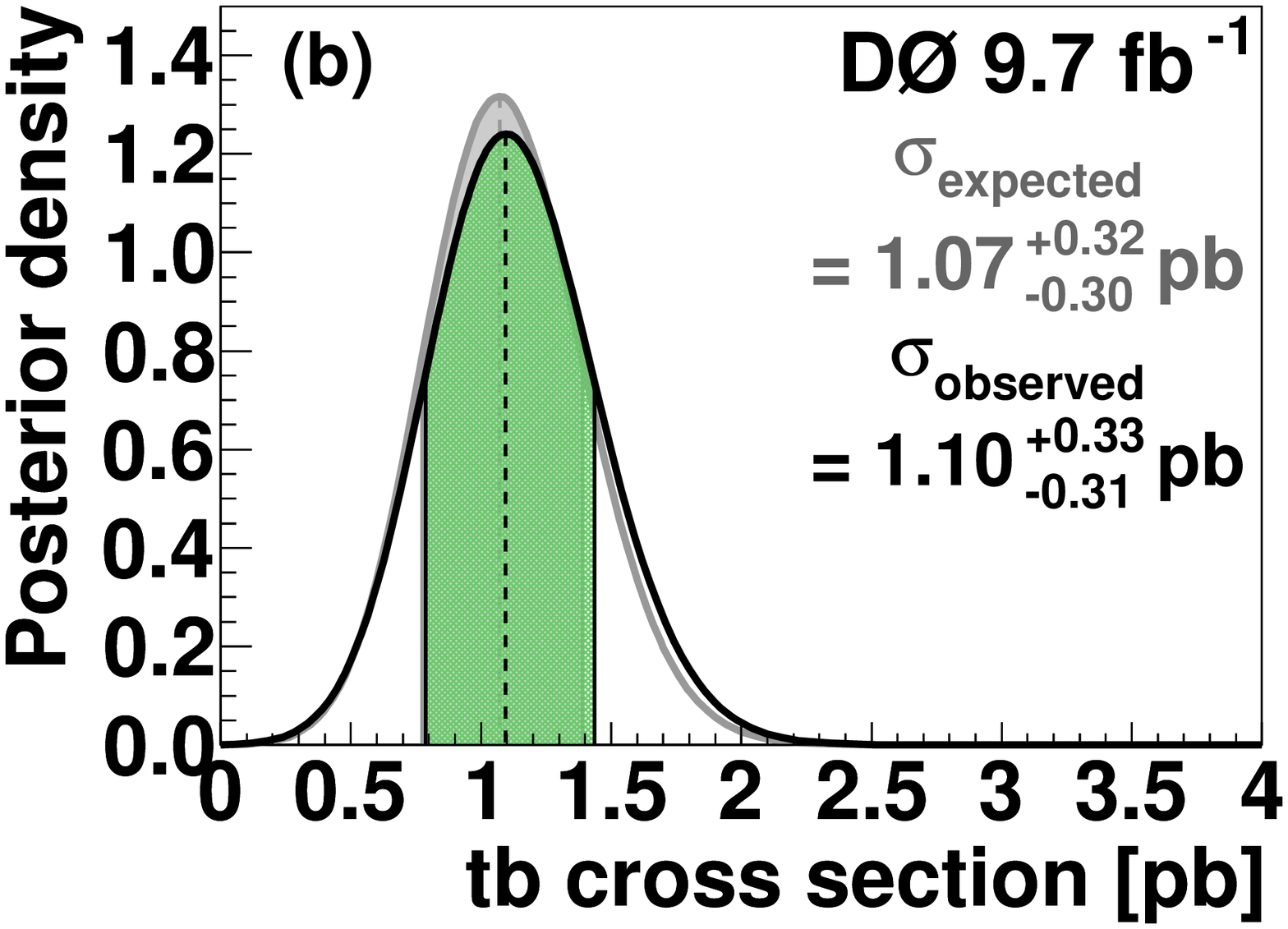}
\includegraphics[width=0.49\textwidth]{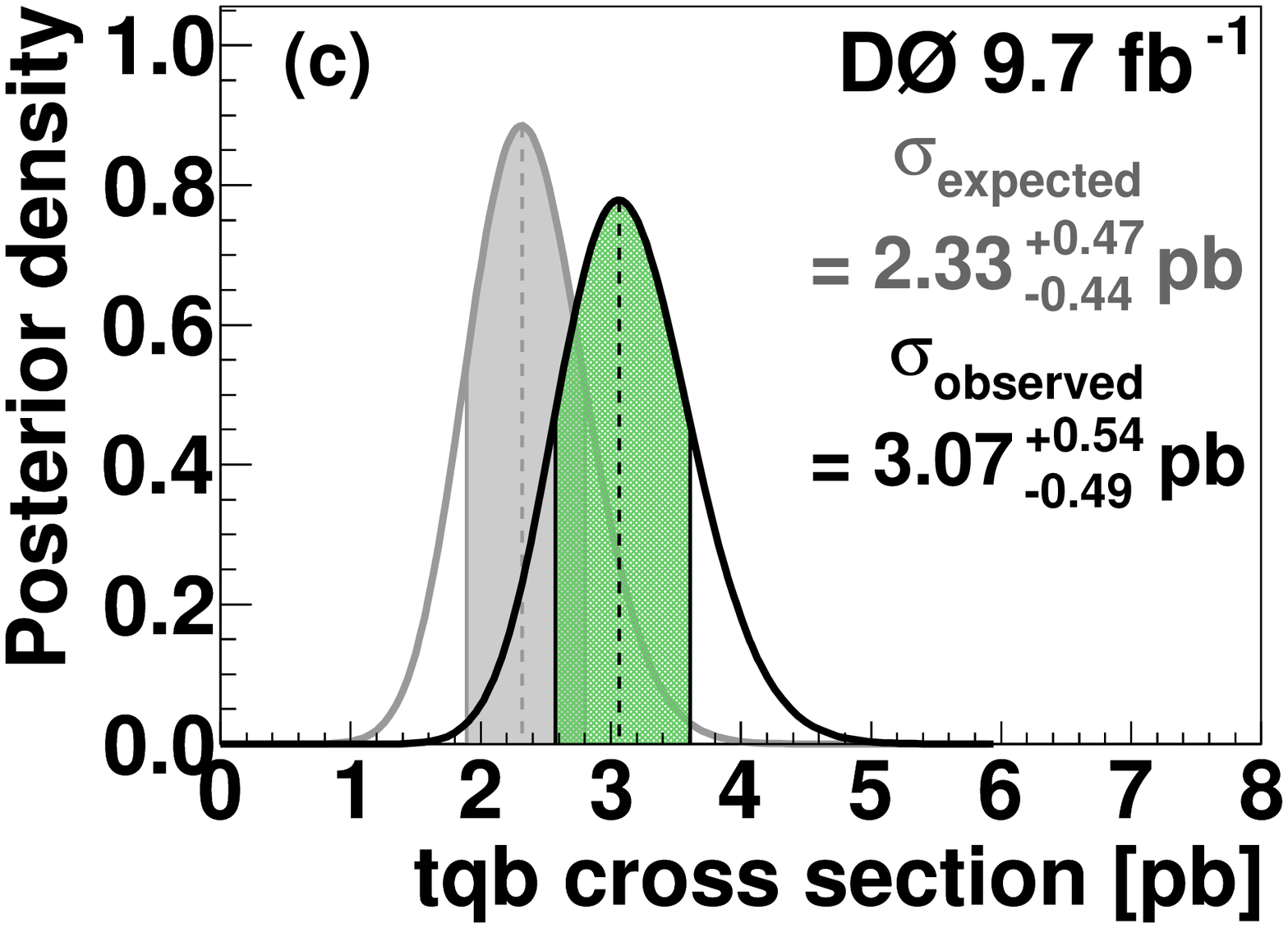}
\includegraphics[width=0.49\textwidth]{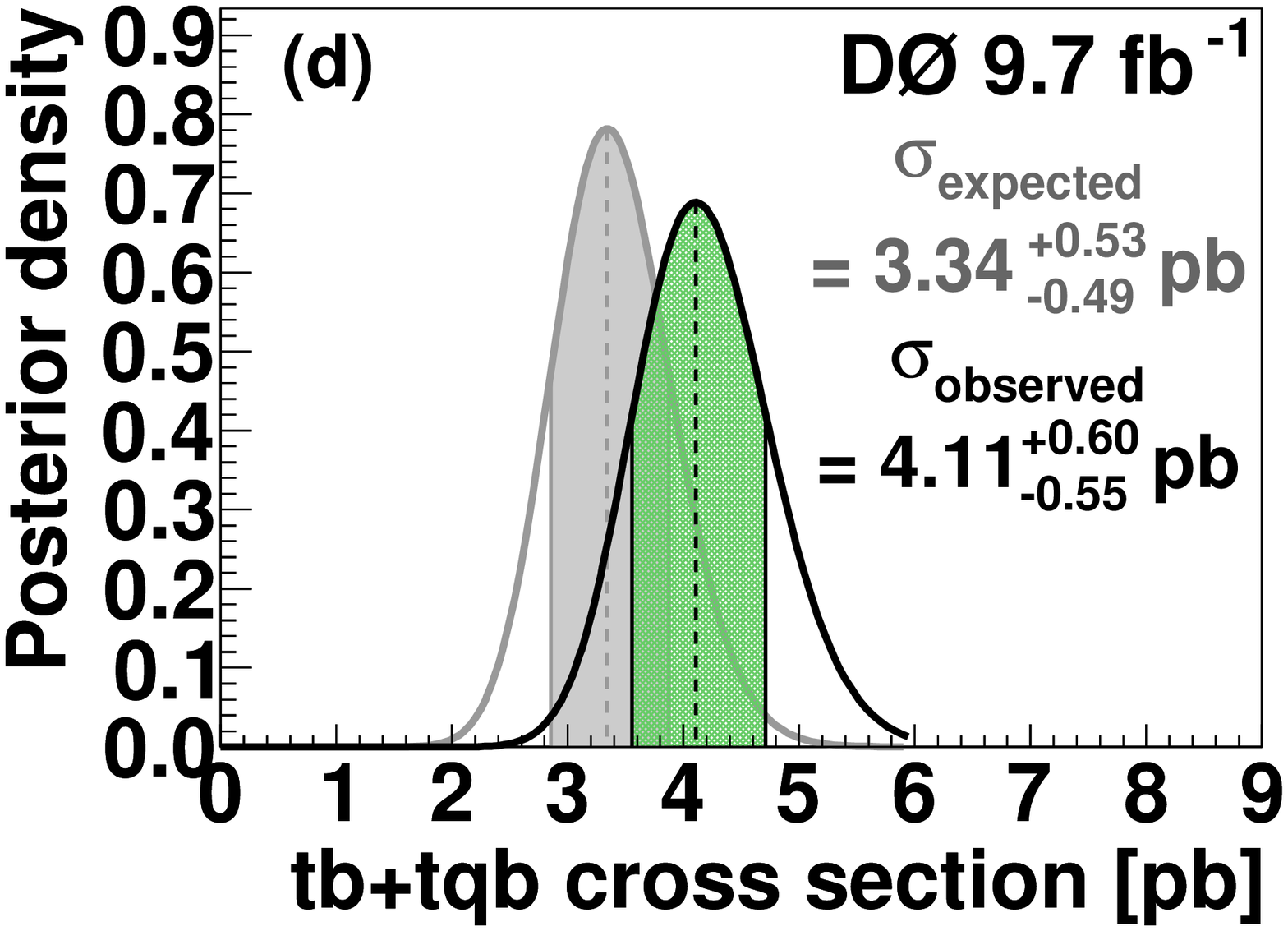}
\caption{(Color online.) Observed posterior density distributions extracted from
the $D_s^{\rm comb}$ and $D_t^{\rm comb}$ discriminants: (a) 2D posterior density with one, two, and three SD probability contours,
and the resulting 1D posterior densities for (b) the $s$ channel,
(c) the $t$ channel, and (d) $s+t$ channel, all along with expected posterior densities.
The prediction from the SM~\cite{singletop-xsec-kidonakis} together with several BSM predictions are shown in (a), including four-quark generations with CKM matrix element $|V_{ts}| = 0.2$~\cite{Alwall:2007}, a top-flavor model with new heavy bosons with $m_x = 1$\;TeV~\cite{Tait:2000sh}, a model of charged top-pions with $m_{\pi^{\pm}} = 250$\;GeV~\cite{Tait:2000sh}, and flavor changing neutral currents with an up-quark/top-quark/gluon coupling $\kappa_u/\Lambda = 0.036$~\cite{Tait:2000sh,d0-fcnc}.}
\label{fig:2d_plots}
\end{center}
\end{figure*}

Figure~\ref{fig:2d_plots} shows contours of equal probabilities for a given number of standard deviations in the 2D posterior for the combined discriminant. The figure also shows the sensitivity to some models of BSM physics that would change the $s$- or $t$-channel cross sections.
To measure the uncertainty on the individual cross sections, we obtain the one-dimensional (1D) posterior probability functions by integrating the 2D posterior over the other variable. 
To measure the combined $s+t$ cross section $\sigma_{s+t}$ without assuming the SM ratio of $\sigma_s/\sigma_t$,
a 2D posterior of $\sigma_{s+t}$ versus $\sigma_t$ is first formed and then the 1D estimate of $\sigma_{s+t}$ found by integrating over all possible values of $\sigma_t$. The results of these measurements are summarized in Table~\ref{tab:xs_results}. 
\begin{table*}[htdp]
\begin{center}
\caption{The expected and observed single top quark cross sections and $p$ values for the individual ME, BNN, and BDT discriminants, and the combined BNN discriminant $D^{\rm comb}$. Here, $Z$ is defined such that a $Z$ standard-deviation upward fluctuation of a Gaussian random variable would have an upper tail area equal to the $p$ value.}
\tabcolsep=5pt
\begin{tabular}{lcccccc}
\hline
\hline
Channel & Expected $\sigma$ (pb) & Observed $\sigma$ (pb) & Expected $p$ value & Observed $p$ value & Expected $Z$ & Observed $Z$ \\
\hline
ME$_s$           & $1.05^{+0.36}_{-0.34}$ & $1.12^{+0.36}_{-0.33}$ & $8.1 \times 10^{-4}$ & $3.7 \times 10^{-4}$ & 3.2 & 3.4 \\
BNN$_s$          & $1.06^{+0.41}_{-0.39}$ & $1.61^{+0.43}_{-0.40}$ & $3.3 \times 10^{-3}$ & $1.5 \times 10^{-5}$ & 2.7 & 4.2 \\
BDT$_s$          & $1.06^{+0.35}_{-0.33}$ & $1.56^{+0.40}_{-0.37}$ & $5.4 \times 10^{-4}$ & $2.3 \times 10^{-6}$ & 3.3 & 4.6 \\
$D_s^{\rm comb}$ & $1.07^{+0.32}_{-0.30}$ & $1.10^{+0.33}_{-0.31}$ & $1.0 \times 10^{-4}$ & $1.0 \times 10^{-4}$ & 3.7 & 3.7 \\ \hline
ME$_t$           & $2.27^{+0.55}_{-0.51}$ & $2.15^{+0.54}_{-0.50}$ & $6.6 \times 10^{-7}$ & $2.8 \times 10^{-6}$ & 4.8 & 4.5 \\
BNN$_t$          & $2.31^{+0.54}_{-0.50}$ & $2.41^{+0.55}_{-0.51}$ & $2.4 \times 10^{-7}$ & $1.4 \times 10^{-7}$ & 5.0 & 5.1 \\
BDT$_t$          & $2.36^{+0.53}_{-0.50}$ & $3.70^{+0.66}_{-0.60}$ & $5.4 \times 10^{-8}$ & $3.4 \times 10^{-15}$ & 5.3 & 7.8 \\
$D_t^{\rm comb}$ & $2.33^{+0.47}_{-0.44}$ & $3.07^{+0.54}_{-0.49}$ & $1.0 \times 10^{-9}$ & $7.1\times 10^{-15}$ & 6.0 & 7.7 \\ \hline
$D_{s+t}^{\rm comb}$ & $3.34^{+0.53}_{-0.49}$ & $4.11^{+0.60}_{-0.55}$ &  &  &  &  \\
\hline
\hline
\end{tabular}
\label{tab:xs_results}
\end{center}
\end{table*}

All three measurements are in agreement with the SM predictions within the uncertainties~\cite{singletop-xsec-kidonakis}.
The statistical significance of these results is quantified by a $p$ value,
which represents the likelihood that the measured cross section could result from a background-only fluctuation equal to or greater than the observed value, assuming the signal process is absent. 
An asymptotic log-likelihood ratio approach~\cite{asymptLLR} is adopted to quantify the $p$ values with the results summarized in Table~\ref{tab:xs_results}. The $s$-channel cross section, without any assumption on the $t$-channel cross section, is measured with a significance corresponding to $3.7$\;SD, which is also the expected sensitivity of our analysis for this process. This is the first measurement of the $s$-channel cross section at more than $3$\;SD. The $t$-channel cross section is measured with $7.7$\;SD ($6.0$\;SD expected). The relative uncertainty on the $s+t$ cross section measurement is improved by $40\%$ with respect to the previous D0 measurement~\cite{stop-2011-d0}, and is now $14\%$, including both statistical and systematic components. 
The statistical component is dominant: the result without systematic uncertainties has a relative uncertainty of $11\%$. 
The experimental dependence of $\sigma_s$ on the assumed value of $m_t$ is $-0.08$\;pb/GeV, and for $\sigma_t$ is $-0.04$\;pb/GeV.

The single top quark production cross section is directly proportional to the square of the 
CKM matrix element $|V_{tb}|^2$, enabling us to measure $|V_{tb}|$ directly without any assumption on the number of quark families or the unitarity of the CKM matrix~\cite{d0-prd-2008}. 
We assume only SM processes for single top quark production and top quarks to decay exclusively to $Wb$. 
We also assume that the {\sl Wtb} interaction is CP-conserving and of the type $V-A$, but maintain the possibility for an anomalous strength of the left-handed {\sl Wtb} coupling ($f_1^L$), which could rescale the single top quark cross section~\cite{flv}. Therefore, we are measuring the strength of the $V-A$ coupling $|V_{tb}f^L_1|$, which can be greater than 1.

We start from the same combination BNN discriminants for $s$ and $t$ channels, and form a Bayesian posterior probability density for $|V_{tb}f^L_1|^2$ with a flat prior, without any assumption on the $\sigma_s/\sigma_t$ production ratio. Additional theoretical uncertainties are considered for the $s$- and $t$-channel cross sections~\cite{singletop-xsec-kidonakis}. 
We obtain $|V_{tb}f^L_1| = 1.12^{+0.09}_{-0.08}$. If we restrict the 
prior to the SM region [0,1] and assume $f^L_1=1$, we extract a limit of $|V_{tb}| > 0.92$ at $95\%$~C.L. 

In summary, we have measured the single top quark production cross section using the full Run II dataset collected by the D0 experiment at the Fermilab Tevatron Collider, corresponding to an integrated luminosity of 9.7\;fb$^{-1}$ after application of appropriate data quality requirements.
We measure the cross sections for $s$ channel and $t$ channel independently, assuming $m_t=172.5$\;GeV: 
\[\sigma({\ppbar}{\rargap}tb+X) = 1.10^{+0.33}_{-0.31}\;\rm pb,\]
\[\sigma({\ppbar}{\rargap}tqb+X) = 3.07^{+0.54}_{-0.49}\;\rm pb.\]
With no assumption on the relative $s$- and $t$-channel contributions, we measure the total single top quark production cross section to be
\[\sigma({\ppbar}{\rargap}tb+tqb+X) = 4.11^{+0.60}_{-0.55}\;\rm pb.\]
All measurements are consistent with the SM predictions~\cite{singletop-xsec-kidonakis}. The $s$-channel production is measured with a significance of $3.7$\;SD and represents the first evidence for this production mode.
Finally, we derive a direct limit on the CKM matrix element, $|V_{tb}| > 0.92$ at $95\%$~C.L.,
assuming a flat prior within $0 \leq |V_{tb}|^2 \leq 1$.


\input{acknowledgement.tex}


\end{document}

%% file: author_list.tex
\affiliation{LAFEX, Centro Brasileiro de Pesquisas F\'{i}sicas, Rio de Janeiro, Brazil}
\affiliation{Universidade do Estado do Rio de Janeiro, Rio de Janeiro, Brazil}
\affiliation{Universidade Federal do ABC, Santo Andr\'e, Brazil}
\affiliation{University of Science and Technology of China, Hefei, People's Republic of China}
\affiliation{Universidad de los Andes, Bogot\'a, Colombia}
\affiliation{Charles University, Faculty of Mathematics and Physics, Center for Particle Physics, Prague, Czech Republic}
\affiliation{Czech Technical University in Prague, Prague, Czech Republic}
\affiliation{Institute of Physics, Academy of Sciences of the Czech Republic, Prague, Czech Republic}
\affiliation{Universidad San Francisco de Quito, Quito, Ecuador}
\affiliation{LPC, Universit\'e Blaise Pascal, CNRS/IN2P3, Clermont, France}
\affiliation{LPSC, Universit\'e Joseph Fourier Grenoble 1, CNRS/IN2P3, Institut National Polytechnique de Grenoble, Grenoble, France}
\affiliation{CPPM, Aix-Marseille Universit\'e, CNRS/IN2P3, Marseille, France}
\affiliation{LAL, Universit\'e Paris-Sud, CNRS/IN2P3, Orsay, France}
\affiliation{LPNHE, Universit\'es Paris VI and VII, CNRS/IN2P3, Paris, France}
\affiliation{CEA, Irfu, SPP, Saclay, France}
\affiliation{IPHC, Universit\'e de Strasbourg, CNRS/IN2P3, Strasbourg, France}
\affiliation{IPNL, Universit\'e Lyon 1, CNRS/IN2P3, Villeurbanne, France and Universit\'e de Lyon, Lyon, France}
\affiliation{III. Physikalisches Institut A, RWTH Aachen University, Aachen, Germany}
\affiliation{Physikalisches Institut, Universit\"at Freiburg, Freiburg, Germany}
\affiliation{II. Physikalisches Institut, Georg-August-Universit\"at G\"ottingen, G\"ottingen, Germany}
\affiliation{Institut f\"ur Physik, Universit\"at Mainz, Mainz, Germany}
\affiliation{Ludwig-Maximilians-Universit\"at M\"unchen, M\"unchen, Germany}
\affiliation{Panjab University, Chandigarh, India}
\affiliation{Delhi University, Delhi, India}
\affiliation{Tata Institute of Fundamental Research, Mumbai, India}
\affiliation{University College Dublin, Dublin, Ireland}
\affiliation{Korea Detector Laboratory, Korea University, Seoul, Republic of Korea}
\affiliation{CINVESTAV, Mexico City, Mexico}
\affiliation{Nikhef, Science Park, Amsterdam, The Netherlands}
\affiliation{Radboud University Nijmegen, Nijmegen, The Netherlands}
\affiliation{Joint Institute for Nuclear Research, Dubna, Russia}
\affiliation{Institute for Theoretical and Experimental Physics, Moscow, Russia}
\affiliation{Moscow State University, Moscow, Russia}
\affiliation{Institute for High Energy Physics, Protvino, Russia}
\affiliation{Petersburg Nuclear Physics Institute, St. Petersburg, Russia}
\affiliation{Instituci\'{o} Catalana de Recerca i Estudis Avan\c{c}ats (ICREA) and Institut de F\'{i}sica d'Altes Energies (IFAE), Barcelona, Spain}
\affiliation{Uppsala University, Uppsala, Sweden}
\affiliation{Lancaster University, Lancaster LA1 4YB, United Kingdom}
\affiliation{Imperial College London, London SW7 2AZ, United Kingdom}
\affiliation{The University of Manchester, Manchester M13 9PL, United Kingdom}
\affiliation{University of Arizona, Tucson, Arizona 85721, USA}
\affiliation{University of California Riverside, Riverside, California 92521, USA}
\affiliation{Florida State University, Tallahassee, Florida 32306, USA}
\affiliation{Fermi National Accelerator Laboratory, Batavia, Illinois 60510, USA}
\affiliation{University of Illinois at Chicago, Chicago, Illinois 60607, USA}
\affiliation{Northern Illinois University, DeKalb, Illinois 60115, USA}
\affiliation{Northwestern University, Evanston, Illinois 60208, USA}
\affiliation{Indiana University, Bloomington, Indiana 47405, USA}
\affiliation{Purdue University Calumet, Hammond, Indiana 46323, USA}
\affiliation{University of Notre Dame, Notre Dame, Indiana 46556, USA}
\affiliation{Iowa State University, Ames, Iowa 50011, USA}
\affiliation{University of Kansas, Lawrence, Kansas 66045, USA}
\affiliation{Louisiana Tech University, Ruston, Louisiana 71272, USA}
\affiliation{Northeastern University, Boston, Massachusetts 02115, USA}
\affiliation{University of Michigan, Ann Arbor, Michigan 48109, USA}
\affiliation{Michigan State University, East Lansing, Michigan 48824, USA}
\affiliation{University of Mississippi, University, Mississippi 38677, USA}
\affiliation{University of Nebraska, Lincoln, Nebraska 68588, USA}
\affiliation{Rutgers University, Piscataway, New Jersey 08855, USA}
\affiliation{Princeton University, Princeton, New Jersey 08544, USA}
\affiliation{State University of New York, Buffalo, New York 14260, USA}
\affiliation{University of Rochester, Rochester, New York 14627, USA}
\affiliation{State University of New York, Stony Brook, New York 11794, USA}
\affiliation{Brookhaven National Laboratory, Upton, New York 11973, USA}
\affiliation{Langston University, Langston, Oklahoma 73050, USA}
\affiliation{University of Oklahoma, Norman, Oklahoma 73019, USA}
\affiliation{Oklahoma State University, Stillwater, Oklahoma 74078, USA}
\affiliation{Brown University, Providence, Rhode Island 02912, USA}
\affiliation{University of Texas, Arlington, Texas 76019, USA}
\affiliation{Southern Methodist University, Dallas, Texas 75275, USA}
\affiliation{Rice University, Houston, Texas 77005, USA}
\affiliation{University of Virginia, Charlottesville, Virginia 22904, USA}
\affiliation{University of Washington, Seattle, Washington 98195, USA}
\author{V.M.~Abazov} \affiliation{Joint Institute for Nuclear Research, Dubna, Russia}
\author{B.~Abbott} \affiliation{University of Oklahoma, Norman, Oklahoma 73019, USA}
\author{B.S.~Acharya} \affiliation{Tata Institute of Fundamental Research, Mumbai, India}
\author{M.~Adams} \affiliation{University of Illinois at Chicago, Chicago, Illinois 60607, USA}
\author{T.~Adams} \affiliation{Florida State University, Tallahassee, Florida 32306, USA}
\author{J.P.~Agnew} \affiliation{The University of Manchester, Manchester M13 9PL, United Kingdom}
\author{G.D.~Alexeev} \affiliation{Joint Institute for Nuclear Research, Dubna, Russia}
\author{G.~Alkhazov} \affiliation{Petersburg Nuclear Physics Institute, St. Petersburg, Russia}
\author{A.~Alton$^{a}$} \affiliation{University of Michigan, Ann Arbor, Michigan 48109, USA}
\author{A.~Askew} \affiliation{Florida State University, Tallahassee, Florida 32306, USA}
% ...
%
%
\collaboration{The D0 Collaboration\footnote{with visitors from
$^{a}$Augustana College, Sioux Falls, SD, USA,
$^{b}$The University of Liverpool, Liverpool, UK,
$^{c}$DESY, Hamburg, Germany,
$^{d}$Universidad Michoacana de San Nicolas de Hidalgo, Morelia, Mexico
$^{e}$SLAC, Menlo Park, CA, USA,
$^{f}$University College London, London, UK,
$^{g}$Centro de Investigacion en Computacion - IPN, Mexico City, Mexico,
$^{h}$Universidade Estadual Paulista, S\~ao Paulo, Brazil,
$^{i}$Karlsruher Institut f\"ur Technologie (KIT) - Steinbuch Centre for Computing (SCC)
and
$^{j}$Office of Science, U.S. Department of Energy, Washington, D.C. 20585, USA.
}} \noaffiliation
\vskip 0.25cm

%% file: acknowledgement.tex
%
We thank the staffs at Fermilab and collaborating institutions,
and acknowledge support from the
DOE and NSF (USA);
CEA and CNRS/IN2P3 (France);
MON, NRC KI and RFBR (Russia);
CNPq, FAPERJ, FAPESP and FUNDUNESP (Brazil);
DAE and DST (India);
Colciencias (Colombia);
CONACyT (Mexico);
NRF (Korea);
FOM (The Netherlands);
STFC and the Royal Society (United Kingdom);
MSMT and GACR (Czech Republic);
BMBF and DFG (Germany);
SFI (Ireland);
The Swedish Research Council (Sweden);
and
CAS and CNSF (China).
%

%% file: plb.bbl
\begin{thebibliography}{99}
\bibitem{eta}
The pseudorapidity is defined as $\eta = -\ln[\tan(\theta/2)]$,
where $\theta$ is the polar angle relative to the proton beam
direction.
\bibitem{EC}
Forward electrons ($1.5 < |\eta| < 2.5$) are not used in this analysis
due to their lower reconstruction efficiency and higher backgrounds.
\end{thebibliography}
